\documentclass[english,keywords,showpacs,amsmath,amstext,amssymb]{iopart}
\usepackage{graphicx}
\usepackage{epstopdf}
\usepackage{epsfig}
\usepackage{mathrsfs}
\usepackage{amssymb}
\usepackage{iopams}
\usepackage{upgreek}
\usepackage{bbold}
\usepackage{color}
\usepackage{bm}
\usepackage{longtable}

\begin{document}

\title{Quantum coherence dynamics of displaced squeezed thermal state in a Non-Markovian environment}

\author{Md.~Manirul Ali\footnote{manirul@citchennai.net}$^1$, 
R. Chandrashekar \footnote{chandrashekar@physics.iitm.ac.in}$^{2,1}$ and S.S. Naina Mohammed \footnote{naina@gacudpt.in (corresponding author)}$^{3}$}
\address{$^1$ Centre for Quantum Science and Technology, Chennai Institute of Technology, Chennai 600069, India}
\address{$^2$ Centre for Quantum Information, Communication and Computing, Indian Institute of Technology Madras, Chennai 600036, India}
\address{$^3$ Department  of Physics, Government Arts College, Udumalpet, Tiruppur District  - 642126, Tamilnadu, India}

\begin{abstract}
The dynamical behavior of quantum coherence of a displaced squeezed thermal state in contact with an external bath is discussed in the 
present work.  We use a Fano-Anderson type of Hamiltonian to model the environment and solve the quantum Langevin equation.  From 
the solution of the quantum Langevin equation we obtain the Green's functions which are used to calculate the expectation value of the 
quadrature operators which are in turn used to construct the covariance matrix.  We use a relative entropy based measure to calculate the
quantum coherence of the mode.  The single mode squeezed thermal state is studied in the Ohmic, sub-Ohmic and the super-Ohmic 
limits for different values of the mean photon number.  In all these limits, we find that when the coupling between the system and 
the environment is weak, the coherence decays monotonically and exhibit a Markovian nature.  When the system and the environment 
are strongly coupled, we observe that the evolution is initially Markovian and after some time it becomes non-Markovian.  The non-Markovian 
effect is due to the environmental back action on the system. Finally, we also present the steady state dynamics of the coherence in the long time 
limit in both low and high temperature regime.  We find that the qualitative behaviour remains the same in both the low and high temperature limits.
But quantitative values differ because the coherence in the system is lower due to thermal decoherence.  
\end{abstract}


\maketitle

\section{Introduction}
Quantum information has evolved from being a theoretical field to technological one in the last few decades \cite{georgescu2012quantum}.  
In most of the theoretical investigations, the quantum systems are considered as isolated objects where the effect of the external environment 
is not included in the study. However in actual situations, the quantum system is exposed to an external environment which can 
cause decoherence in the system \cite{scully1997quantum}. In general, we model the environment as a many body quantum system which 
is also referred to as bath. When the quantum system is coupled to the bath, there is a dynamical change in the quantum properties 
of the system. Both the qualitative and quantitative nature of this dynamical change depends on the bath properties as well as on the 
coupling between the system and the environment. Qualitatively, there are two fundamental kinds of dynamics namely 
the Markovian and the non-Markovian dynamics \cite{breuer2016colloquium,de2017dynamics}. A knowledge of the dynamical change is 
important from the point of view of fabrication of quantum devices  
\cite{vyas2020master,astafiev2006temperature,ribeiro2015non,yoshihara2006decoherence}.
When the relaxation time of the bath is very short compared to the evolution time of the system, the quantumness of the 
system falls monotonically which is a Markovian decay \cite{davies1974markovian,davies1976markovian,gorini1978properties}.  
For the non-Markovian dynamics, the bath relaxation time and the system evolution time are comparable to each other 
\cite{zhang2012general,de2017dynamics}. Due to this we will observe a revival of quantumness due to environmental back action
\cite{franco2012revival,robinett2004quantum}. Thus an open quantum system has a very rich and interesting structure which has formed 
the basis for several seminal works.  

Continuous variable systems can describe the interaction and propagation of electromagnetic waves and hence are a 
valuable resource in quantum information processing \cite{braunstein2005quantum,weedbrook2012gaussian,schumaker1986quantum}.  
An electromagnetic field with quantized radiation modes can be denoted by bosonic modes. For $n$ number of bosonic modes, 
the  Hilbert space is  $\prod_{k=1}^{n} H_{k}$.  The creation and annihilation operators of the $k^{th}$ bosonic field is given by  
$a^{\dag}_{k}$ and $a_{k}$ respectively.  Alternatively, the continuous variable system can be described using the 
quadrature operators $\{x_{k}, p_{k} \}$ and for a $n$-mode system we have a $2n$-dimensional vector which contains 
all the quadrature pairs.  For a single mode continuous variable system, the quadrature operators are 
$\xi_{1} = x = (a + a^{\dag})$ and $\xi_{2} = p = -i (a - a^{\dag})$ and the $2D$ quadrature vector $\bm{\xi} = \{ \xi_{1}, \xi_{2} \}$. 
It is well known that the quadrature operators satisfy the canonical commutation relation  $[ \xi_{i}, \xi_{j} ] = 2i \Omega_{ij}$,
with $\Omega_{ij}$ being the elements of the matrix
\begin{equation}
\bm{\Omega} = \left(\begin{array}{cc}
 0  & 1\\
-1 & 0\\
\end{array}
\right).
\end{equation}
A continuous variable quantum system which has representation in terms of Gaussian functions is referred to as a Gaussian state. 
Theoretically Gaussian states are easier to investigate and also experimentally they are easier to produce and hence several discussions 
on continuous variable states are restricted to Gaussian states.  In particular, a Gaussian state is completely characterized by the 
first and second moments of the quadrature field operators.  For these operators we can construct a vector of the first moments 
$\overline{\bm{\xi}} = ( \langle \xi_{1} \rangle, \langle \xi_{2} \rangle )$ and the covariance matrix $\bm{V}$
\begin{equation}
V_{ij} = \langle \{ \Delta \xi_i, \Delta \xi_j \} \rangle = {\rm Tr} \left( \{ \Delta \xi_i, \Delta \xi_j \} \rho \right). 
\end{equation}
Here we consider $\{ \Delta \xi_i, \Delta \xi_j \} = (\Delta \xi_i \Delta \xi_j + \Delta \xi_j \Delta \xi_i )/2$,  and the 
fluctuation operator is $\Delta \xi_{i}  =  \xi - \langle \xi_{i} \rangle$.  In the present work we consider a quantum system which is a 
single mode Gaussian state for which the matrix elements of the covariance matrix are 
\begin{eqnarray}
V_{ii} &=& \langle \xi_i^2 \rangle - \langle \xi_i \rangle^2, \\
V_{ij} &=& \frac{1}{2} \langle \xi_i \xi_j + \xi_j \xi_i \rangle - \langle \xi_i \rangle \langle \xi_j \rangle.
\label{covele}
\end{eqnarray}
Due to the positivity of the density matrix the covariance matrix has to satisfy the uncertainty relation $\bm{V} + i \bm{\Omega} \ge 0$. 

Entanglement is one of the fundamental properties of a quantum system and has been studied in detail for finite 
dimensional and infinite dimensional systems.  But entanglement is not the only unique property, rather it is one among an 
hierarchy  of quantum properties.  In the hierarchy, quantum coherence occupies the topmost level which nestles within it the other 
properties like nonlocality, steering, entanglement and discord \cite{ma2019operational}.  This means that in a given 
quantum system, even when these properties are not present, quantum coherence might be present in the system.  
A scheme to estimate quantum coherence was introduced by Baumgratz et al., \cite{baumgratz2014quantifying} from the perspective 
of quantum information theory.  This led to an explosion of interest in the field of quantum coherence especially in defining new quantum 
coherence measures \cite{radhakrishnan2016distribution}, resource theory of quantum coherence \cite{streltsov2017colloquium,chitambar2019quantum} 
and also in some applications \cite{hillery2016coherence,zhang2019demonstrating,ma2019operational1}.  
Initially, most of these works were done on quantum systems with finite degrees of freedom. The quantum coherence of the 
finite dimensional system \cite{radhakrishnan2016distribution} is measured as the relative entropy distance between 
two matrices 
\begin{equation}
\mathcal{D}(\rho) =  \min_{\sigma} S(\rho \| \sigma) = \min_{\sigma} {\rm Tr} (\rho \log_{2} \rho - \rho \log_{2} \sigma),
\label{coher1}
\end{equation}
where $\rho$ and $\sigma$ being the given quantum state and the incoherent state respectively.  For a continuous variable 
system, the quantum coherence was initially investigated using a density matrix approach \cite{zhang2016quantifying}.  
This method did not give a closed form expression for Gaussian states.  Hence in Ref. \cite{xu2016quantifying}, 
the authors made use of a covariance matrix method to quantify coherence in the system since the covariance matrix 
can completely characterizes a Gaussian state.  The coherence measure based on the covariance matrix is 
\begin{equation}
C(\rho) = \inf_{\delta}  S(\rho \| \delta).
\end{equation}
Here the minimization runs over all the incoherent Gaussian states. The closest incoherent state to a Gaussian state is 
the thermal state of the form
\begin{equation}
\rho_d = \sum_{n=0}^{\infty} \frac{\overline{\mu}^n}{(1+\overline{\mu})^{n+1}} \vert n \rangle \langle n \vert,
\label{thermal}
\end{equation}
with $\overline{\mu}={\rm Tr} [a^\dagger a \rho]$ being the mean photon number associated with the Gaussian state. The entropy of a single mode
Gaussian state is 
\begin{equation}
S(\rho) = \frac{\nu+1}{2} \log_{2}  \frac{\nu+1}{2} - \frac{\nu-1}{2} \log_{2}  \frac{\nu-1}{2},
\end{equation}
where $\nu = \sqrt{{\rm det} {\bm V}}$. Hence the relative entropy of coherence for the single mode Gaussian state is 
\begin{eqnarray}
C(\rho) &=& S(\rho \| \rho_d) ={ \rm Tr} (\rho \log_2 \rho - \rho \log_2 \rho_d) \nonumber \\
&=& \frac{\nu-1}{2} \log_2 \frac{\nu-1}{2} - \frac{\nu+1}{2} \log_2 \frac{\nu+1}{2} \nonumber \\
& & + (\overline{\mu}+1) \log_2 (\overline{\mu}+1) - \overline{\mu} \log_2 \overline{\mu}.
\label{coher3}
\end{eqnarray}
For our work, we use this relative entropy measure to estimate the coherence in the single mode quantum system \cite{rivas2010entanglement}. 
Investigations on the open system dynamics of entanglement has been carried out on both qubit (finite dimensional) systems
as well as on infinite dimensional (continuous variable) systems \cite{Paz2008dynamics,Zhang2007hong,Liu2007Goan}.  
The entanglement dynamics presents unique features like
sudden death \cite{yu2009sudden} and revival of entanglement \cite{mazzola2009sudden}.  In the case of quantum coherence the 
dynamics has been investigated only for qubit systems \cite{Pati2018Banerjee,chandra2019time,radhakrishnan2019dynamics}.  
In this work, we investigate the dynamics of a single mode Gaussian 
state which is a displaced squeezed thermal state.  In Sec. \ref{formulation}, of the article we give a description of the system 
and the environment, and also the procedure to calculate the time evolved covariance matrix.  We consider three different spectral 
densities in this manuscript and in Sec. \ref{ohmic}, we compute the quantum coherence in the Ohmic environment.  
In Sec. \ref{subohmic} and Sec. \ref{superohmic} we describe the dynamical behavior of coherence in the sub-Ohmic and 
the super-Ohmic environments.  A steady state analysis of the system is given in Sec. \ref{steadystate} and we give our 
conclusions in Sec. \ref{conclusions}.

\section{Formulation of the system-environment model and the dynamics}
\label{formulation}
The quantum coherence dynamics of finite dimensional systems have been studied from both the theoretical 
\cite{chandra2019time,radhakrishnan2019dynamics} and experimental \cite{cao2020fragility}
perspectives.  An infinite dimensional case is considered in the present work where we consider the system to be a single 
bosonic mode of frequency $\omega_{0}$.  The bath coupled to the system is a non-Markovian environment at a finite 
temperature.  This non-Markovian environment is a structured bosonic reservoir \cite{lambropoulos2000fundamental}
with a collection of infinite modes of varying frequencies.  The system-environment combination can be described using the Fano-Anderson Hamiltonian 
\cite{anderson1961localized,fano1961effects}
\begin{equation}
H = \hbar \omega_0 a^{\dagger}a + \hbar \sum_k \omega_k b_k^{\dagger}b_k
      + \hbar \sum_k  \left(\mathcal{V}_k a^{\dagger} b_k  + \mathcal{V}_k^{\ast} b_k^{\dagger}a \right).
\label{Tot}
\end{equation}
Here, the factor  $\mathcal{V}_{k}$ represents the coupling strength between the bath and the system and 
$a^{\dag}$ ($a$) is the creation (annihilation) operator where $\omega_{0}$ is the frequency of the system.  
For the $k^{th}$ mode of the bosonic reservoir with frequency $\omega_{k}$,  $b^{\dag}_{k}$ ($b_{k}$) 
is the corresponding creation (annihilation) operator.  This Hamiltonian is used in the study of several different 
models in the fields of atomic and condensed matter physics.  

To solve the dynamics, we can use the Heisenberg equation of motion approach.  The time evolved operators 
corresponding to the system and the environment are $a(t) = e^{\frac{i H t}{\hbar}} a e^{-\frac{i H t}{\hbar}}$
and $b_k(t) = e^{\frac{i H t}{\hbar}} b_k e^{-\frac{i H t}{\hbar}}$ and in the Heisenberg picture they satisfy
\begin{eqnarray}
&  & \frac{d}{dt} a(t) =  -\frac{i}{\hbar}  \left[ a(t), H \right] = -i \omega_0 a(t) - i \sum_k \mathcal{V}_k b_k(t),  \quad
\label{EOM1} \\
&  & \frac{d}{dt} b_k(t) = - \frac{i}{\hbar} \left[ b_k(t), H \right] = -i \omega_k b_k(t) - i \mathcal{V}_k^{\ast} a(t).   \quad
\label{EOM2}
\end{eqnarray}
To obtain the quantum Langevin equation \cite{ali2017nonequilibrium} we solve Eqn. (\ref{EOM2}) for $b_{k}$ and substitute the result in 
Eqn. (\ref{EOM1}) which gives 
\begin{equation}
{\dot a}(t) + i \omega_0 a(t)  + \int_0^t  d\tau g(t,\tau) a(\tau) = - i \sum_k \mathcal{V}_k b_k(0) e^{-i\omega_k t}.
\label{langevinequation}
\end{equation}
The non-Markovian memory effects between the system and the environment is characterized by the integral kernel 
$g(t,\tau)  = \sum_k |\mathcal{V}_k|^2 e^{-i\omega_k (t-\tau)}$.  In the case of an environment with a continuous 
spectrum $g(t,\tau) = \int_0^{\infty} d\omega J(\omega) e^{-i\omega(t-\tau)}$ with 
$J(\omega)=\varrho(\omega) |\mathcal{V}(\omega)|^2$ being the spectral density characterizing the non-Markovian 
memory of the environment.  The factor $\varrho(\omega)$ is the density of states of the environment and $\omega$
is the continuously varying bath frequency.   Since the quantum Langevin equation in Eqn. (\ref{langevinequation}) 
is linear, one can write $a(t) = u(t) a(0) + f(t)$.  Here the time dependent coefficients $u(t)$ and the noise 
operator satisfy the two integro-differential equations given below:
\begin{eqnarray}
\frac{d}{dt} u(t) &=&   - i \omega_0 u(t) - \int_0^t d\tau g(t,\tau) u(\tau), 
\label{utime} \\
\frac{d}{dt} f(t)  &=&   -i \omega_0 f(t)  -  \int_0^t d\tau g(t,\tau) f(\tau)  \nonumber \\
                            &   & - i \sum_k \mathcal{V}_k b_k(0) e^{-i\omega_k t}.   
\label{gtime}
\end{eqnarray}
We can find $u(t)$ by numerically solving Eqn. (\ref{utime}) with the initial condition $u(0)=1$.  To get $f(t)$ we solve 
Eq. (\ref{gtime}) subject to the initial condition $f(0)=0$ which gives
\begin{equation}
f(t) = - i \sum_k \mathcal{V}_k b_k(0) \int_0^t d\tau e^{-i\omega_k \tau} u(t,\tau).
\label{ft2}
\end{equation}
The nonequilibrium thermal fluctuation is characterized by the correlation function given below
\begin{equation}
\langle f^{\dagger}(t) f(t)\rangle = v(t) = \! \! \int_{0}^{t}  \! \! \! d\tau_1   \int_{0}^{t^\prime} \! \! \!   d\tau_2 u(t,  \tau_1)
{\widetilde g}(\tau_1,\tau_2) u^{\ast}(t^{\prime},  \tau_2). 
\label{fdf}
\end{equation}
Here we consider the initial state of the total system to be an uncorrelated product state i.e.,
 $\rho_{tot}(0)=\rho_S(0) \otimes \rho_E(0)$. For the environment with the Hamiltonian 
 $H_E=\sum_k \hbar \omega_k b_k^{\dagger}b_k$, where $\beta = 1/k_{B} T$ is the inverse
temperature and $k_B$ is the Boltzmann constant, the thermal state $\rho_E(0) = \exp(-\beta H_E)/{\rm Tr}[\exp(-\beta H_E)]$
is its initial environment state. 

Here the time correlation function of the environment with continuous spectrum is 
\begin{equation}
{\widetilde g}(\tau_1,\tau_2)  = \int_0^{\infty} d\omega  J(\omega) {\bar n}(\omega) e^{-i\omega(\tau_1-\tau_2)}.
\end{equation}
Here ${\bar n}(\omega)=1/(e^{\hbar \omega / k_{B} T}-1)$ is the initial particle number distribution.  The time dependent average
values of the system are: 
\begin{eqnarray}
\label{avg1}
&\langle a(t) \rangle = u(t) \langle a(0) \rangle, ~~\langle a^{\dagger}(t) \rangle = u^{\ast}(t) \langle a^{\dagger}(0) \rangle,  \\
\label{avg2}
&\langle a(t) a(t) \rangle = (u(t))^{2}  \langle a(0) a(0) \rangle, \\
\label{avg3}
&\langle a^{\dagger}(t) a^{\dagger}(t)\rangle = (u^{\ast}(t))^{2}  \langle a^{\dagger}(0) a^{\dagger}(0) \rangle, \\
\label{avg4}
&\langle a^{\dagger}(t) a(t) \rangle =  |u(t)|^{2}  \langle a^{\dagger}(0) a(0) \rangle + v(t).
\end{eqnarray}

\begin{figure}
\includegraphics[width=\columnwidth]{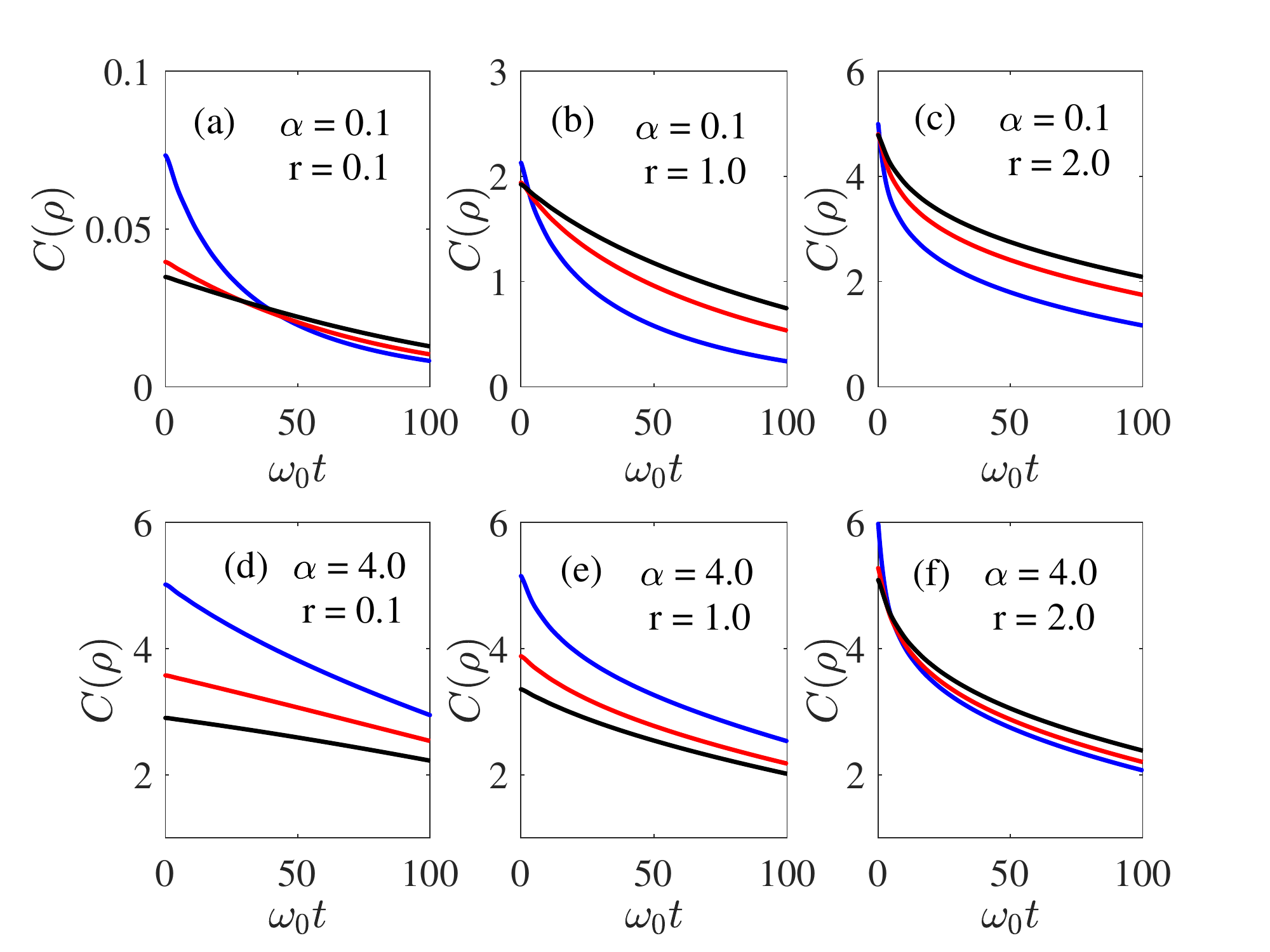}
\caption{The time evolution of quantum coherence of a thermal state in the low temperature limit $T_{s} =1$, is shown above 
for a weakly coupled system ($\eta = 0.01 \; \eta_{c}$) for various values of the displacement parameter (`$\alpha$') and 
squeezing parameters (`$r$').  The different lines correspond to the different values of $\bar{n}$ as follows:  
$\bar{n} = 0.1$ (blue), $\bar{n} = 1.0$ (red) and $\bar{n} = 10.0$ (black). We use Ohmic spectral density ($s=1$) with the cut-off frequency $\omega_{c} = 5.0 \; \omega_{0}$.}
\label{fig1}
\end{figure}

Initially, the reservoir is in a thermal state and uncorrelated to the system so $\langle f(t)\rangle=\langle f^{\dagger}(t)\rangle=0$ 
and also $\langle f(t) f(t) \rangle=\langle f^{\dagger}(t) f^{\dagger}(t)\rangle=0$.  The time evolved first and second moments
of the quadrature operators {\it viz} $\langle \xi_1(t) \rangle$, $\langle \xi_2(t) \rangle$, $\langle \xi_1^2(t) \rangle$,
$\langle \xi_2^2(t) \rangle$, $\langle \xi_1(t) \xi_2(t) \rangle$, and $\langle \xi_2(t) \xi_1(t) \rangle$ can be found from the 
time-dependent average values in Eq. (\ref{avg1} - \ref{avg4}).  From the moments of the quadrature operator one can express
the time evolved covariance matrix as 
\begin{eqnarray}
V_{11} &=& 1+ 2 v(t) + 2  |u(t)|^{2} \,  {\rm Cov}(a^{\dagger}(0), a(0))  \nonumber  \\
                   &   &  + (u(t))^{2} \, {\rm Var}(a(0)) + (u^{\ast}(t))^{2} \, {\rm Var}(a^{\dagger}(0)), \qquad
\label{V11}
\end{eqnarray}
\begin{eqnarray}
V_{22} &=& 1+ 2 v(t) + 2  |u(t)|^{2} \,  {\rm Cov}(a^{\dagger}(0), a(0))  \nonumber  \\
                   &   &  - (u(t))^{2} \, {\rm Var}(a(0)) - (u^{\ast}(t))^{2} \, {\rm Var}(a^{\dagger}(0)), \qquad
\label{V22}
\end{eqnarray}
\begin{equation}
V_{12}  =  i  (u^{\ast}(t))^{2} \; {\rm Var}(a^{\dag} (0))  - i ((u(t))^{2}  \,  {\rm Var}(a(0)).
\end{equation}
where ${\rm Cov}(a,b) = \langle a b \rangle - \langle a \rangle \langle b \rangle$ and ${\rm Var}(a) = {\rm Cov}(a,a)$.  
Due to the symmetry of the covariance matrix we also have $V_{12} = V_{21}$.  

From the knowledge of the initial state and the environmental parameters, we can find the time evolved covariance 
matrix elements using the nonequilibrium Green's functions $u(t)$ and $v(t)$.  The spectral density $J(\omega)$ of 
the environment needs to be specified to calculate the Green's function. An Ohmic type spectral density, 
\begin{equation}
J(\omega) = \eta ~\omega \left( \frac{\omega}{\omega_c} \right)^{s-1} ~e^{-\omega/\omega_c},
\label{sd}
\end{equation}
is considered in our work, since this can simulate a large class of thermal baths.  In the above equation, $\omega_{c}$ is the cut-off frequency of the environmental spectra and $\eta$ is the system-bath coupling 
strength. At the critical value of the coupling strength $\eta_{c} = \omega_{0} /(\omega_{c} \Gamma(s))$ a localized 
mode is generated and here $\Gamma(s)$ is the gamma function. The environment is classified as Ohmic for $s=1$
super-Ohmic for $s > 1$ and sub-Ohmic for $s<1$. The dynamics of coherence is measured for the displaced squeezed 
thermal state over a wide range of parameters. In the present work we consider the Hamiltonian which is bilinear in the 
creation and annihilation operators. Hence the Gaussian states preserve their form and remain Gaussian. Throughout 
our work we use a scaled temperature $T_{s} = k_{B} T/ \hbar \omega_{0}$, where $\omega_{0}$ is the frequency of the system. 

\begin{figure}
\includegraphics[width=\columnwidth]{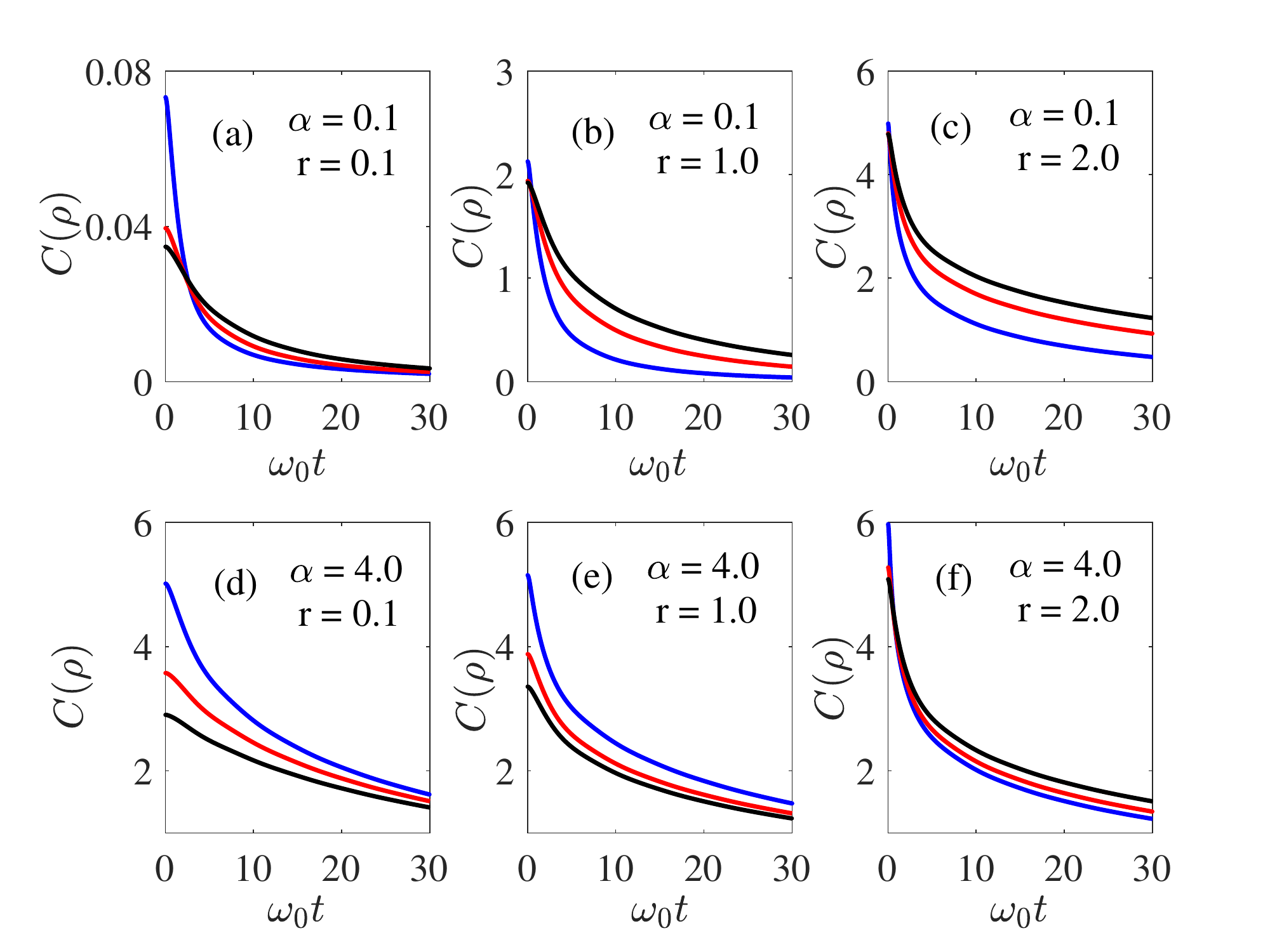}
\caption{The time evolution of quantum coherence of a thermal state in the high temperature limit $T_{s} =20$, is shown above 
for a weakly coupled system ($\eta = 0.01 \; \eta_{c}$) for various values of the displacement parameter (`$\alpha$') and 
squeezing parameters (`$r$').  The different lines correspond to the different values of $\bar{n}$ as follows:  
$\bar{n} = 0.1$ (blue), $\bar{n} = 1.0$ (red) and $\bar{n} = 10.0$ (black). We use Ohmic spectral density ($s=1$) with the cut-off frequency $\omega_{c} = 5.0 \; \omega_{0}$. }
\label{fig2}
\end{figure}

In the present work we investigate the quantum coherence dynamics of a general Gaussian state of the form 
\cite{adam1995density}
\begin{equation}
\rho = D(\alpha) \ S(r) \  \rho_{th} \ S^{\dag}(r) \  D^{\dag}(\alpha),
\label{generalgaussianstate}
\end{equation}
where $D(\alpha)$ and $S(r)$ are the displacement and squeezing operators defined as: 
\begin{eqnarray}
D(\alpha) &=&  \exp(\alpha a^{\dag} - \alpha^{*} a),
\label{displacementoperator}  \\
S(r) &=& \exp \left[ r (a^2 - a^{\dag 2})/2 \right],
\label{squeezingoperator}
\end{eqnarray}
with the thermal state $\rho_{th}$ being: 
\begin{equation}
\rho_{th}  = \sum_{n =0}^{\infty}   \frac{\bar{n}^{n}}{(1+\bar{n})^{n+1}}  |n \rangle \langle n |.
\end{equation}
The transient dynamics of the single mode Gaussian state described above is investigated for different values of the displacement $\alpha$ and squeezing parameter $r$.

\begin{figure}
\includegraphics[width=\columnwidth]{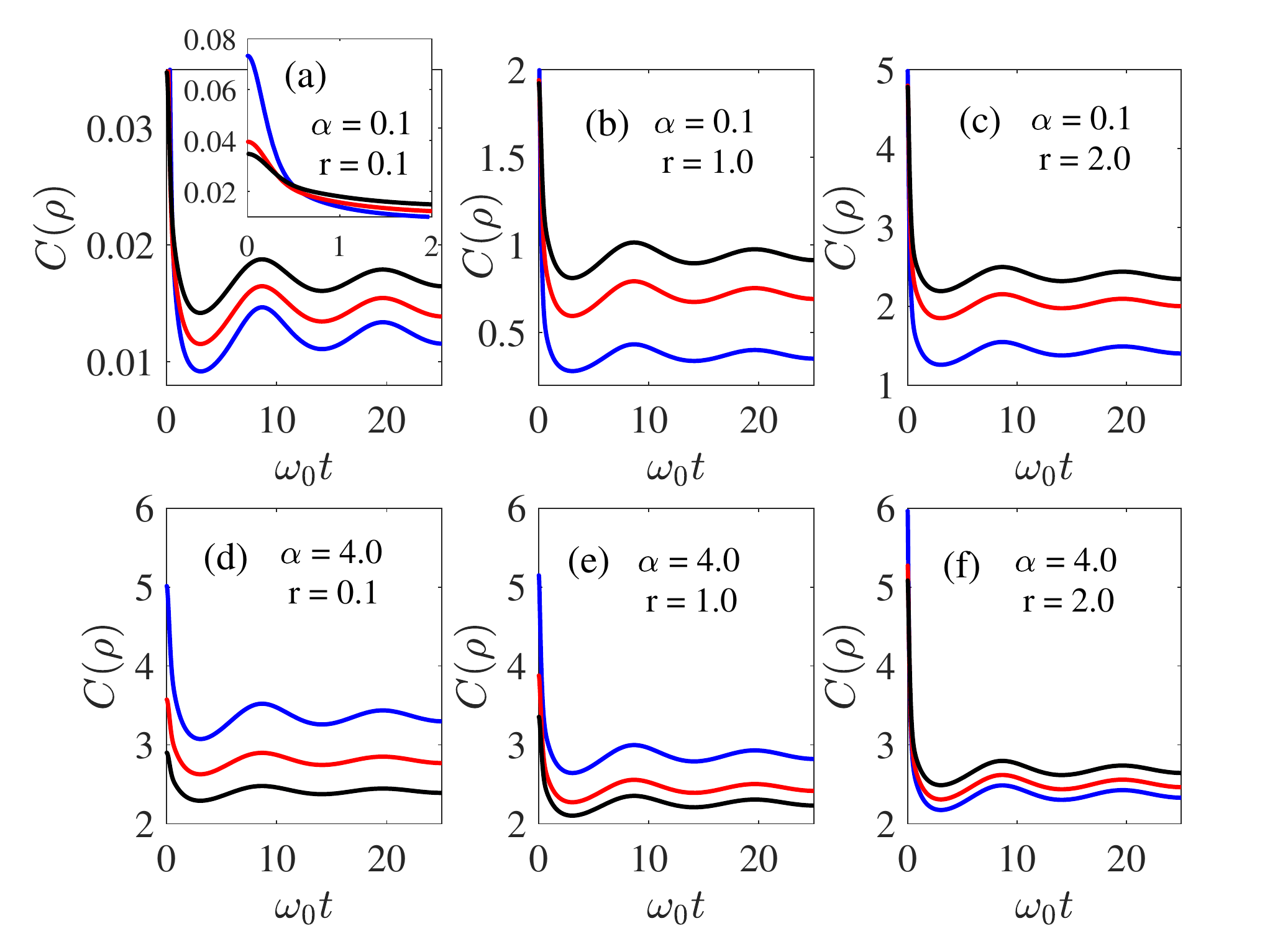}
\caption{The time evolution of quantum coherence of a thermal state in the low temperature limit $T_{s} =1$, is shown above 
for a strongly coupled system ($\eta = 2.0 \; \eta_{c}$) for various values of the displacement parameter (`$\alpha$') and 
squeezing parameters (`$r$').  The different lines correspond to the different values of $\bar{n}$ as follows:  
$\bar{n} = 0.1$ (blue), $\bar{n} = 1.0$ (red) and $\bar{n} = 10.0$ (black). We use Ohmic spectral density ($s=1$) with the cut-off frequency $\omega_{c} = 5.0 \; \omega_{0}$. }
\label{fig3}
\end{figure}

\section{Quantum Coherence evolution of displaced squeezed thermal state in a Ohmic environment}
\label{ohmic}
\begin{figure}
\includegraphics[width=\columnwidth]{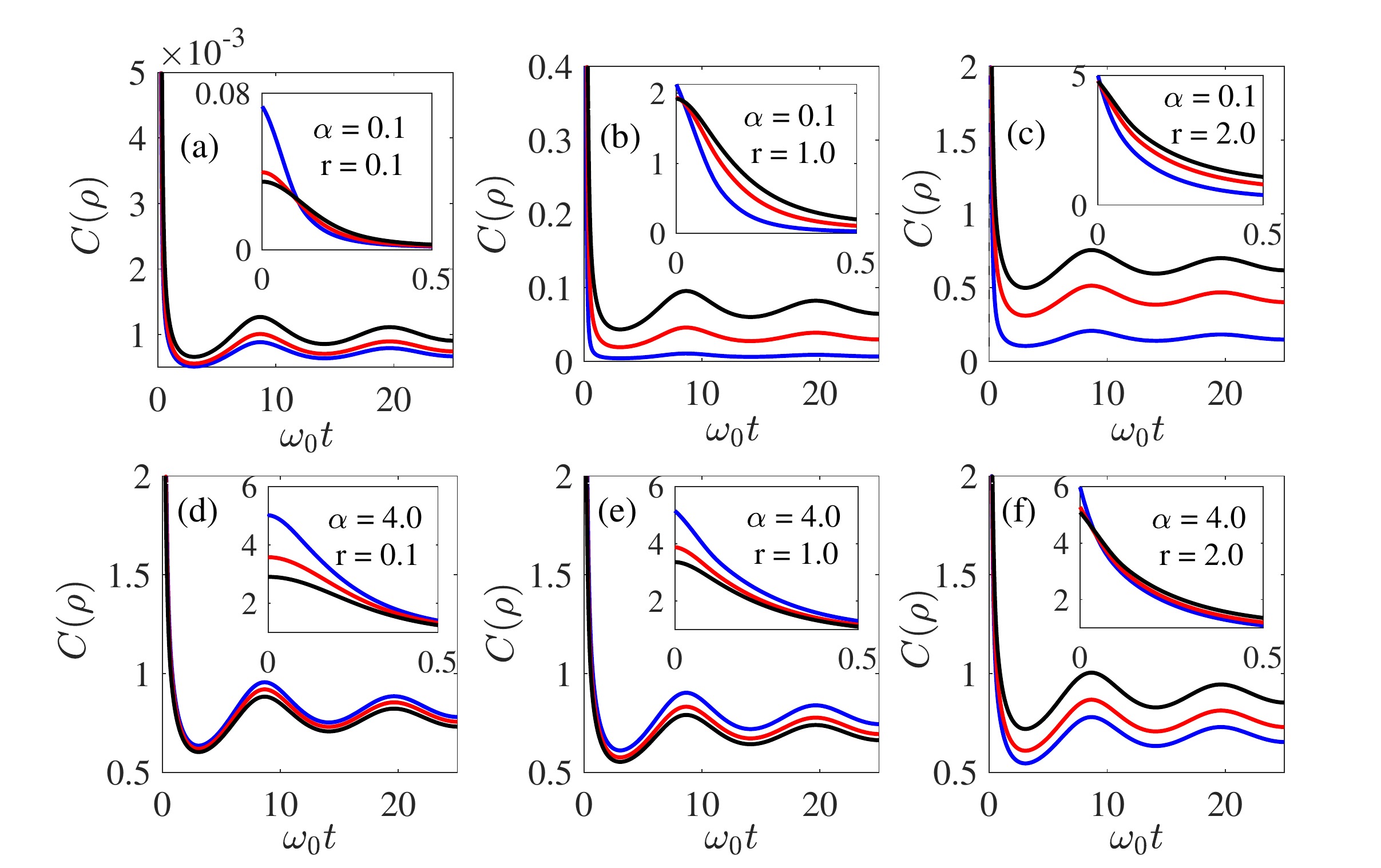}
\caption{The time evolution of quantum coherence of a thermal state in the high temperature limit $T_{s} =20$, is shown above 
for a strongly coupled system ($\eta = 2.0 \; \eta_{c}$) for various values of the displacement parameter (`$\alpha$') and 
squeezing parameters (`$r$').  The different lines correspond to the different values of $\bar{n}$ as follows:  
$\bar{n} = 0.1$ (blue), $\bar{n} = 1.0$ (red) and $\bar{n} = 10.0$ (black). We use Ohmic spectral density ($s=1$) with the cut-off frequency $\omega_{c} = 5.0 \; \omega_{0}$.}
\label{fig4}
\end{figure}

The transient dynamics of a single mode squeezed displaced thermal state in contact with a ohmic bath with spectral 
density $J(\omega) = \eta \ \omega \exp(- \omega/ \omega_{c})$ is described in the present section.  The mode is characterized 
by two parameters {\it viz} the displacement parameter ($\alpha$) and the squeezing parameter ($r$) and the dynamics are 
shown through the plots Fig. \ref{fig1} - Fig. \ref{fig4}. From these plots we find that the amount of initial coherence is 
higher in systems with lower mean photon number.

In Fig. \ref{fig1} and \ref{fig2}, the dynamics of quantum coherence of a weakly coupled system ($\eta = 0.01 \eta_{c}$) is studied
in the low temperature ($T_{s}=1$) and high temperature ($T_{s}=20$) limits respectively.  The rate of fall of coherence decreases with 
increase in the displacement parameter and the squeezing parameter. Since there is no back flow of information, the quantum coherence decreases steadily with time. The initial coherence and the rate of fall also depends on the mean photon number $\bar{n}$. This is characteristic of the system that is weakly coupled  to the environment. A comparison between the low and high temperature plots show that the coherence falls faster in the high temperature limit.  This is because, apart from the loss of coherence due to the dissipative interaction with the environment, the system also suffers from decoherence due to the thermal effects. 

The strongly coupled system ($\eta = 2.0 \eta_{c}$) is studied through the plots in Fig. \ref{fig3} and Fig. \ref{fig4}, corresponding 
to the low temperature ($T_{s}=1$) and high temperature ($T_{s}=20$) limits respectively.  We find that the amount of initial coherence 
is higher at low temperature and the rate of fall of coherence is lesser for systems with higher value of displacement and squeezing parameter.  
The quantum coherence in the system initially falls faster and reaches a minimum value.  Then it increases slightly and exhibits an oscillatory 
behavior.  These oscillations indicate an information back flow which is a characteristic feature of a non-Markovian dynamics.  Due to 
thermal decoherence in the high temperature limit, there is a faster fall of quantum coherence.  Thus we find that the system dynamics 
is dependent on the strength of its coupling with the bath.  

\section{Quantum coherence dynamics of displaced squeezed thermal state in a sub-Ohmic environment}
\label{subohmic}
\begin{figure}
\includegraphics[width=\columnwidth]{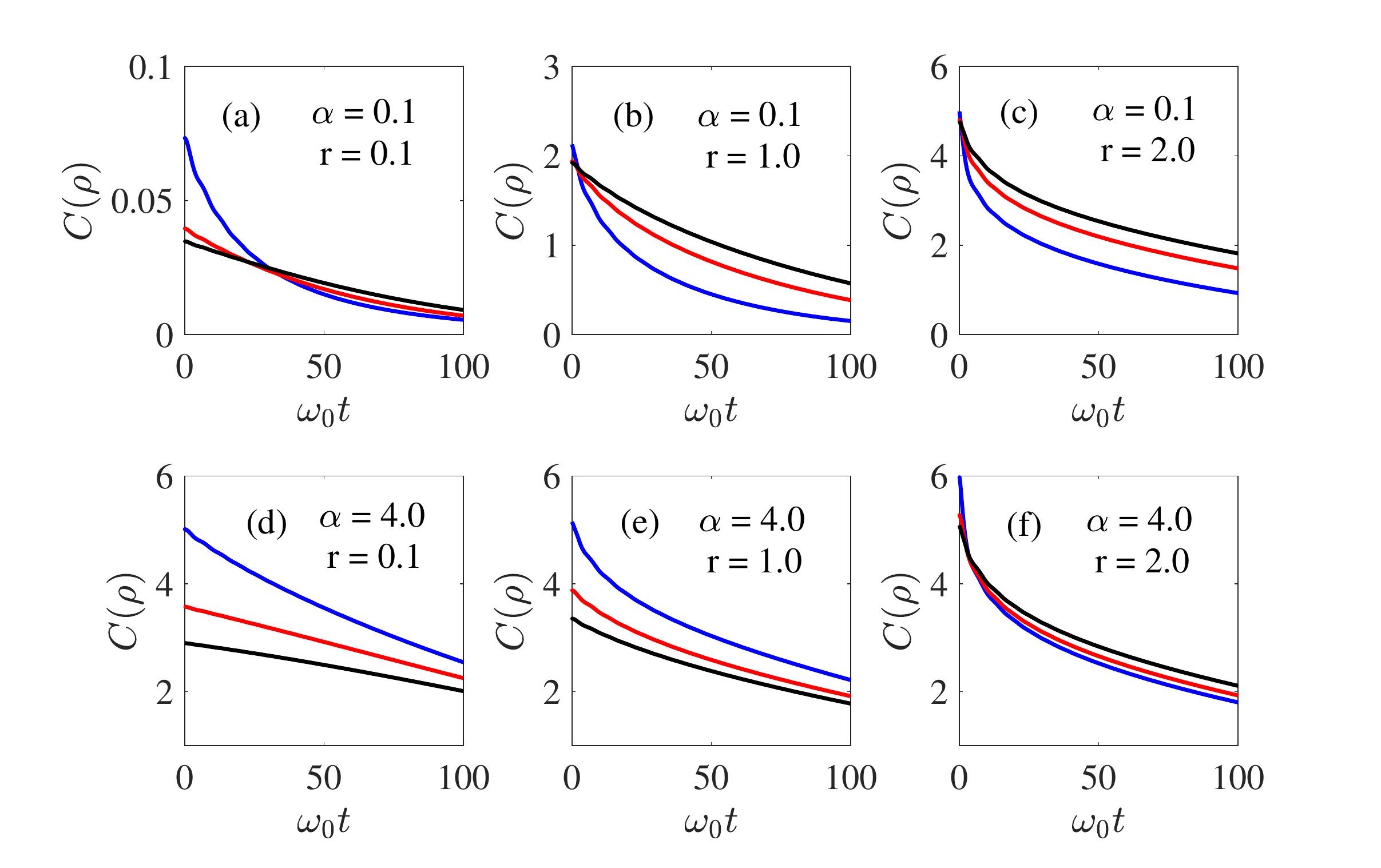}
\caption{The time evolution of quantum coherence of a thermal state in the low temperature limit $T_{s} =1$, is shown above 
for a weakly coupled system ($\eta = 0.01 \; \eta_{c}$) for various values of the displacement parameter (`$\alpha$') and 
squeezing parameters (`$r$').  The different lines correspond to the different values of $\bar{n}$ as follows:  
$\bar{n} = 0.1$ (blue), $\bar{n} = 1.0$ (red) and $\bar{n} = 10.0$ (black). We use sub-Ohmic spectral density ($s=1/2$) with the cut-off frequency $\omega_{c} = 5.0 \; \omega_{0}$.}
\label{fig5}
\end{figure}

\begin{figure}
\includegraphics[width=\columnwidth]{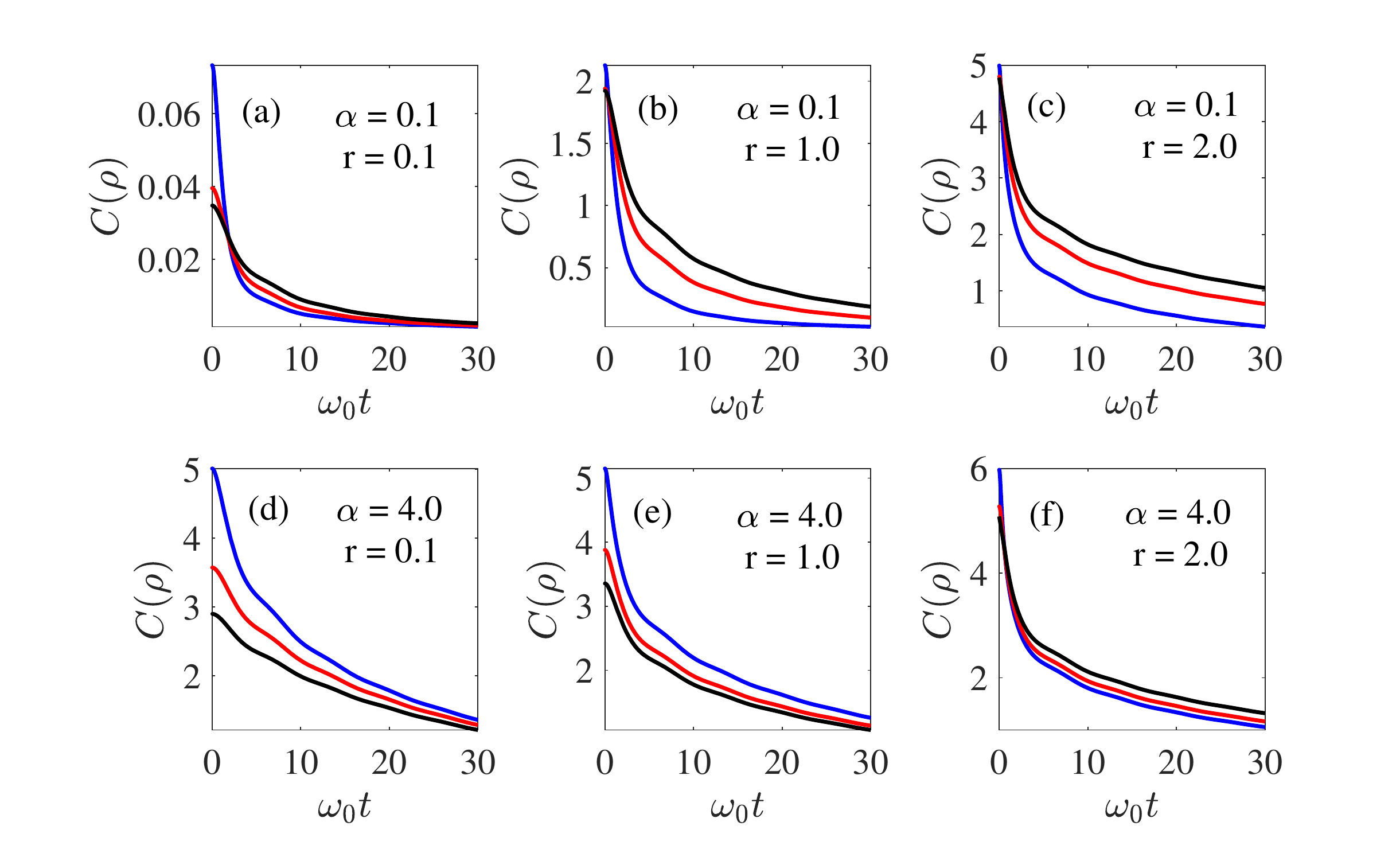}
\caption{The time evolution of quantum coherence of a thermal state in the high temperature limit $T_{s} =20$, is shown above 
for a weakly coupled system ($\eta = 0.01 \; \eta_{c}$) for various values of the displacement parameter (`$\alpha$') and 
squeezing parameters (`$r$').  The different lines correspond to the different values of $\bar{n}$ as follows:  
$\bar{n} = 0.1$ (blue), $\bar{n} = 1.0$ (red) and $\bar{n} = 10.0$ (black). We use sub-Ohmic spectral density ($s=1/2$) with the cut-off frequency $\omega_{c} = 5.0 \; \omega_{0}$. }
\label{fig6}
\end{figure}

\begin{figure}
\includegraphics[width=\columnwidth]{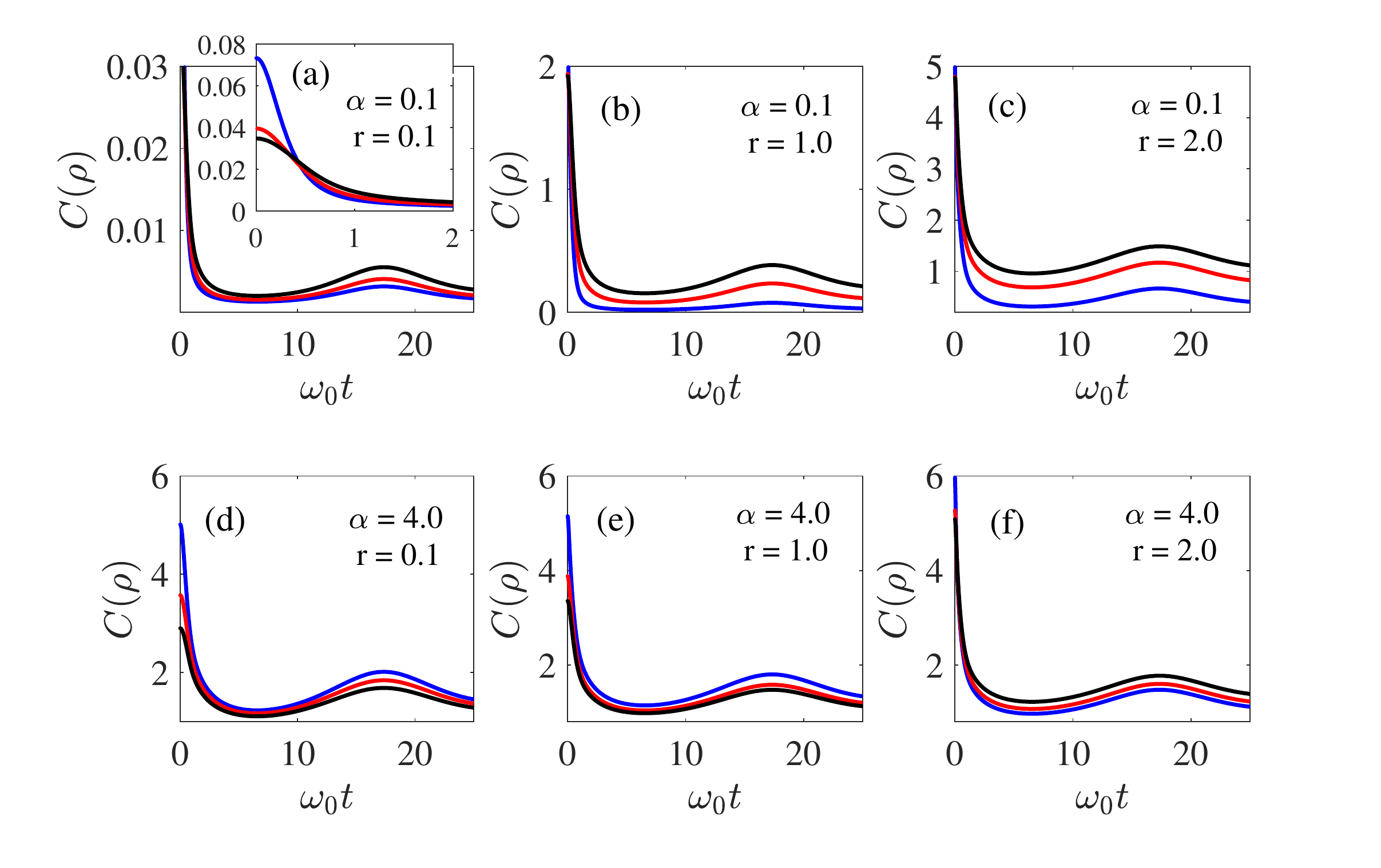}
\caption{The time evolution of quantum coherence of a thermal state in the low temperature limit $T_{s} =1$, is shown above 
for a strongly coupled system ($\eta = 2.0 \; \eta_{c}$) for various values of the displacement parameter (`$\alpha$') and 
squeezing parameters (`$r$').  The different lines correspond to the different values of $\bar{n}$ as follows:  
$\bar{n} = 0.1$ (blue), $\bar{n} = 1.0$ (red) and $\bar{n} = 10.0$ (black). We use sub-Ohmic spectral density ($s=1/2$) with the cut-off frequency $\omega_{c} = 5.0 \; \omega_{0}$. }
\label{fig7}
\end{figure}

\begin{figure}
\includegraphics[width=\columnwidth]{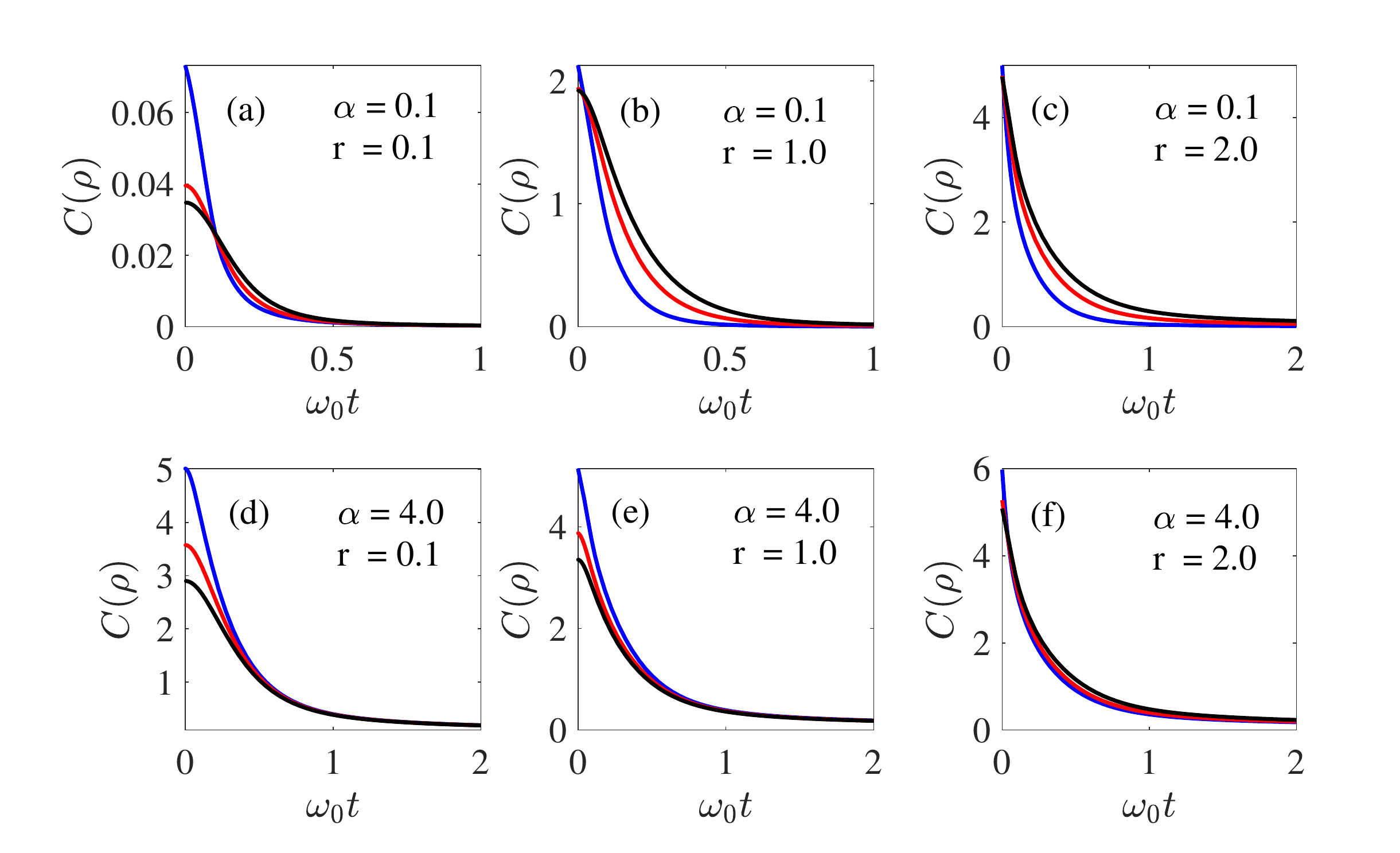}
\caption{The time evolution of quantum coherence of a thermal state in the high temperature limit $T_{s} =20$, is shown above 
for a strongly coupled system ($\eta = 2.0 \; \eta_{c}$) for various values of the displacement parameter (`$\alpha$') and 
squeezing parameters (`$r$').  The different lines correspond to the different values of $\bar{n}$ as follows:  
$\bar{n} = 0.1$ (blue), $\bar{n} = 1.0$ (red) and $\bar{n} = 10.0$ (black). We use sub-Ohmic spectral density ($s=1/2$) with the cut-off frequency $\omega_{c} = 5.0 \; \omega_{0}$.}
\label{fig8}
\end{figure}

A subohmic  spectral density is one in which $s < 1$.  In the present work, we consider $s = 1/2$ and the corresponding spectral density 
reads $J(\omega) = \eta \sqrt{\omega \omega_{c}} \exp(-\omega/\omega_{c})$.  The displacement parameter (`$\alpha$') and the 
squeezing parameter (`$r$') characterizes the mode.  The transient dynamics of quantum coherence of this system under the variation 
of these parameters is shown through the plots in Fig. \ref{fig5} - \ref{fig8}.  

The coherence evolution in the weak coupling limit ($\eta = 0.01 \; \eta_{c}$) is analyzed in Fig. \ref{fig5} and \ref{fig6}, considering the
low temperature and high temperature limits. Here we observe that the amount of initial coherence is inversely proportional to the mean photon number $\bar{n}$. We observe that the amount of initial coherence is inversely proportional to the mean photon number.  In the weak 
coupling limit, the coherence decay exhibits a Markovian nature and decreases steadily with time.  When the displacement parameter 
and the squeezing parameter increases, the coherence decay rate increases.  Also the coherence falls faster in the high temperature 
limit.  This is because thermal decoherence also contributes to the fall of coherence.  

Through the plots Fig. \ref{fig7} and \ref{fig8}, we study the coherence dynamics in a system strongly coupled ($\eta = 2.0 \; \eta_{c}$)
to a sub-ohmic bath.  The low temperature limit is described through the plots in Fig. \ref{fig7} and the high temperature limit 
through Fig. \ref{fig8}.  The amount of initial coherence is higher at lower temperature, with the coherence decay being faster for lower 
values of displacement and squeezing parameter.  Here in the shorter time scales, the coherence falls faster and reaches a minimum 
value.  Then it shows an oscillatory nature indicating a non-Markovian behavior for the strongly coupled system.  This non-Markovian 
dynamics is indicative of information backflow in the system.  Thus we find Markovian evolution for a weakly coupled system and 
a non-Markovian evolution for a strongly coupled system.

\section{Non-Markovian dynamics of displaced squeezed thermal state in a super-Ohmic environment}
\label{superohmic}
\begin{figure}
\includegraphics[width=\columnwidth]{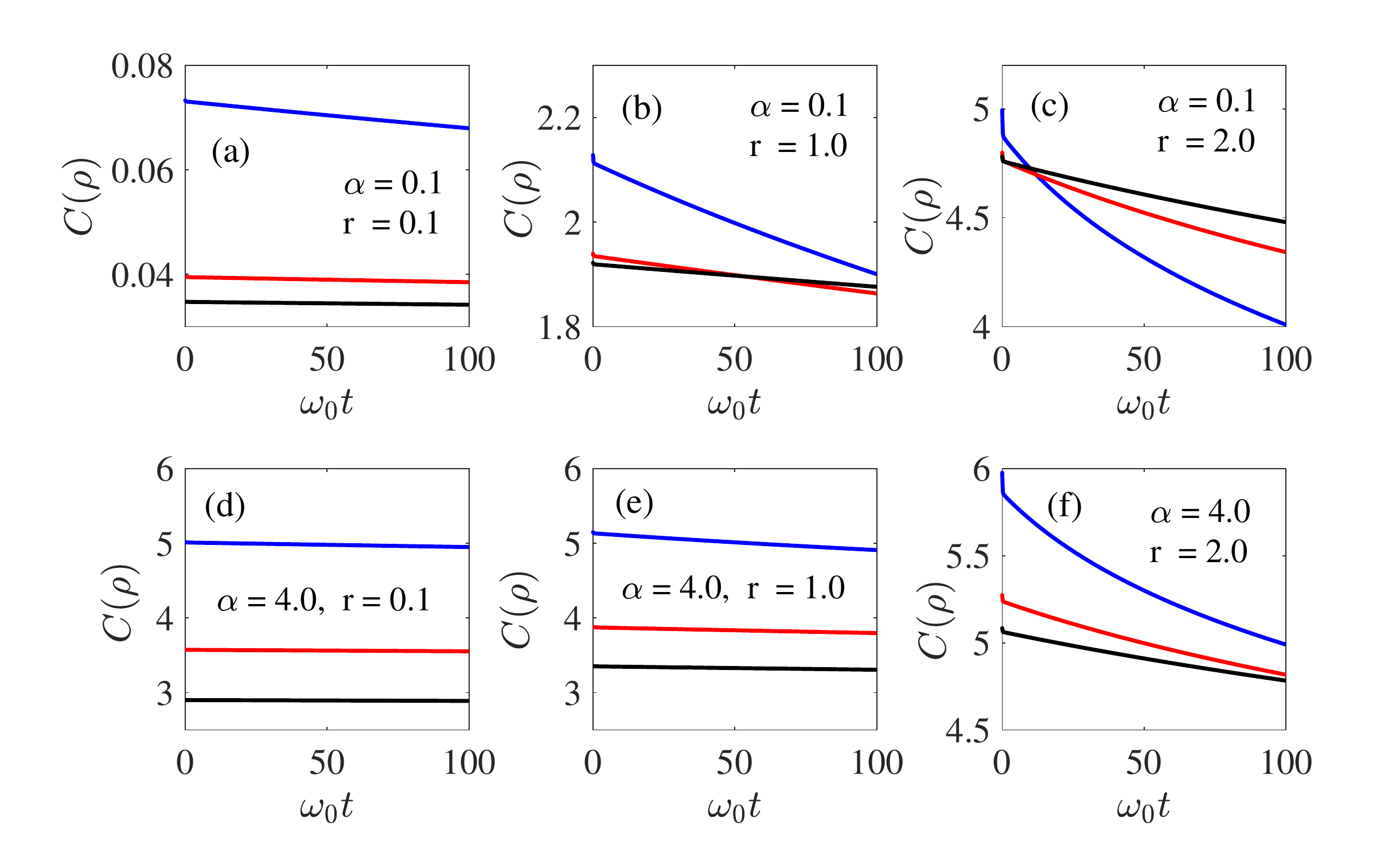}
\caption{The time evolution of quantum coherence of a thermal state in the low temperature limit $T_{s} =1$, is shown above 
for a weakly coupled system ($\eta = 0.01 \; \eta_{c}$) for various values of the displacement parameter (`$\alpha$') and 
squeezing parameters (`$r$').  The different lines correspond to the different values of $\bar{n}$ as follows:  
$\bar{n} = 0.1$ (blue), $\bar{n} = 1.0$ (red) and $\bar{n} = 10.0$ (black). We use super-Ohmic spectral density ($s=3$) with the cut-off frequency $\omega_{c} = 5.0 \; \omega_{0}$.}
\label{fig9}
\end{figure}

\begin{figure}
\includegraphics[width=\columnwidth]{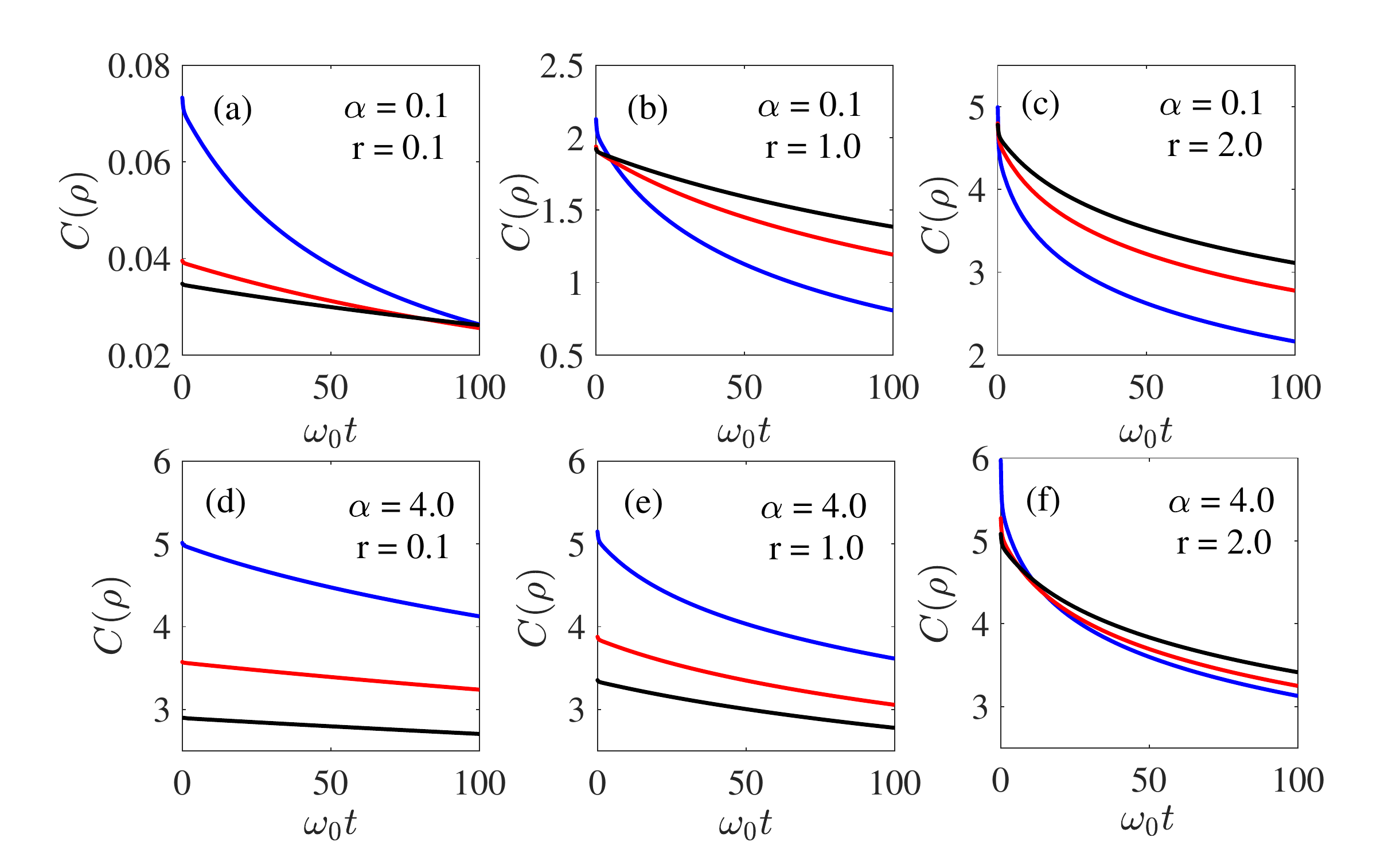}
\caption{The time evolution of quantum coherence of a thermal state in the high temperature limit $T_{s} =20$, is shown above 
for a weakly coupled system ($\eta = 0.01 \; \eta_{c}$) for various values of the displacement parameter (`$\alpha$') and 
squeezing parameters (`$r$').  The different lines correspond to the different values of $\bar{n}$ as follows:  
$\bar{n} = 0.1$ (blue), $\bar{n} = 1.0$ (red) and $\bar{n} = 10.0$ (black). We use super-Ohmic spectral density ($s=3$) with the cut-off frequency $\omega_{c} = 5.0 \; \omega_{0}$. }
\label{fig10}
\end{figure}

\begin{figure}
\includegraphics[width=\columnwidth]{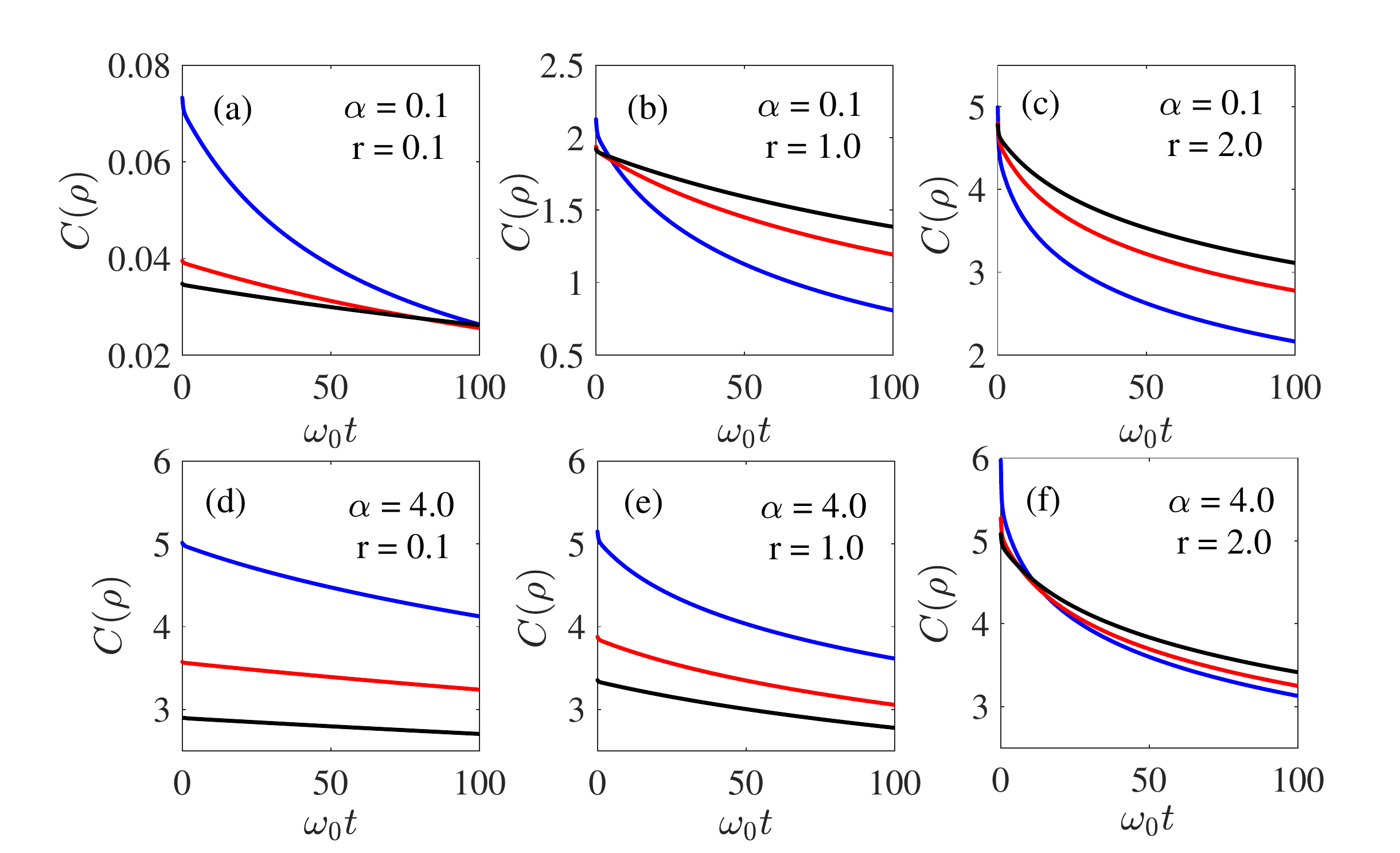}
\caption{The time evolution of quantum coherence of a thermal state in the low temperature limit $T_{s} =1$, is shown above 
for a strongly coupled system ($\eta = 2.0 \; \eta_{c}$) for various values of the displacement parameter (`$\alpha$') and 
squeezing parameters (`$r$').  The different lines correspond to the different values of $\bar{n}$ as follows:  
$\bar{n} = 0.1$ (blue), $\bar{n} = 1.0$ (red) and $\bar{n} = 10.0$ (black). We use super-Ohmic spectral density ($s=3$) with the cut-off frequency $\omega_{c} = 5.0 \; \omega_{0}$. }
\label{fig11}
\end{figure}

\begin{figure}
\includegraphics[width=\columnwidth]{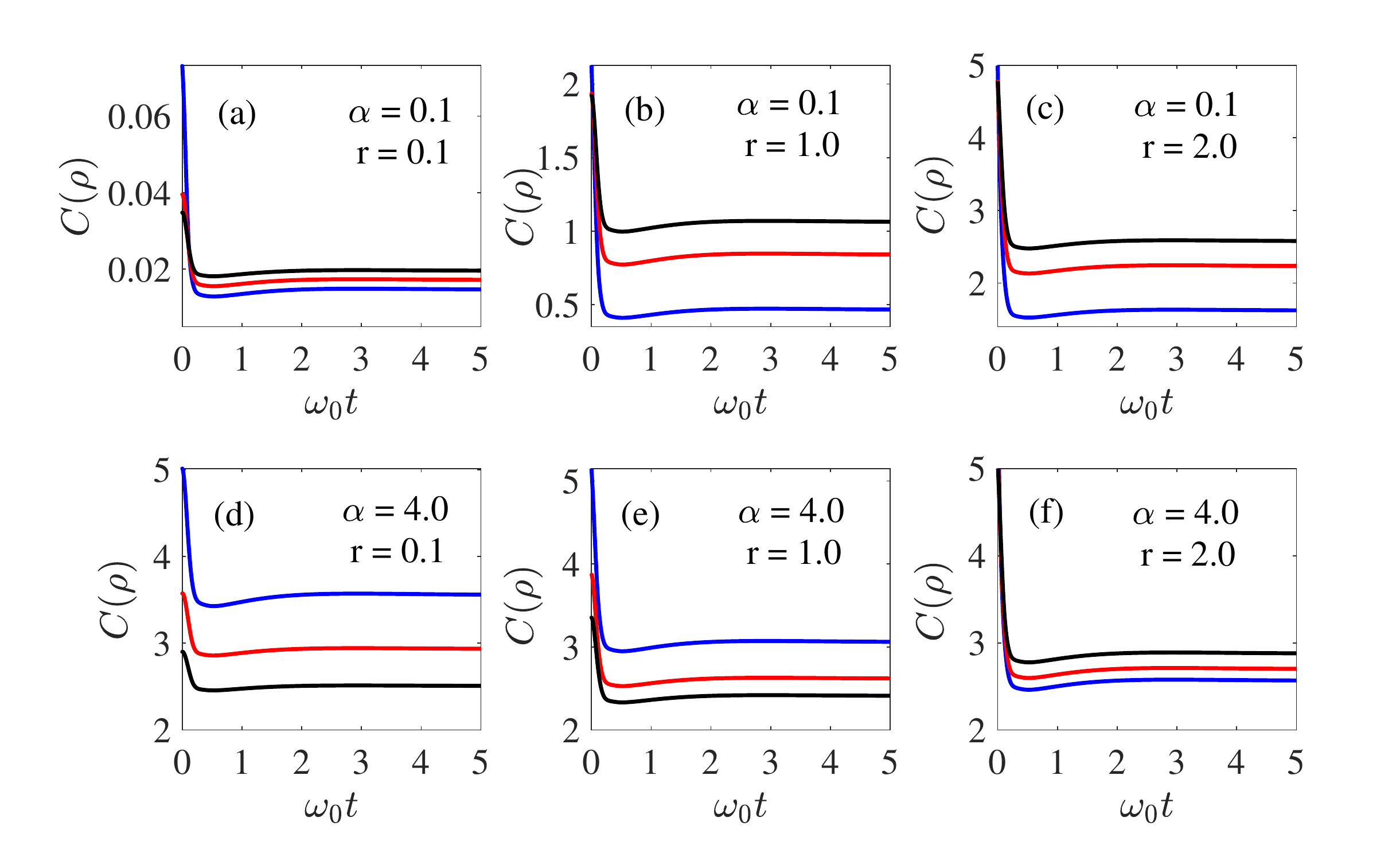}
\caption{The time evolution of quantum coherence of a thermal state in the high temperature limit $T_{s} =20$, is shown above 
for a strongly coupled system ($\eta = 2.0 \; \eta_{c}$) for various values of the displacement parameter (`$\alpha$') and 
squeezing parameters (`$r$').  The different lines correspond to the different values of $\bar{n}$ as follows:  
$\bar{n} = 0.1$ (blue), $\bar{n} = 1.0$ (red) and $\bar{n} = 10.0$ (black). We use super-Ohmic spectral density ($s=3$) with the cut-off frequency $\omega_{c} = 5.0 \; \omega_{0}$.}
\label{fig12}
\end{figure}

The time dynamics of a single mode continuous variable state in contact with a super-Ohmic bath with $s > 1$ is studied in the present work. 
Towards this end, we consider a spectral density of the form $J(\omega) =\eta  (\omega/ \omega_{c})^{3} \exp(-\omega/\omega_{c})$
in our investigations.  The dynamical variation of coherence  is studied by varying the displacement operator and the squeezing parameter. 
The results are displayed in the plots given through Fig. \ref{fig9} - \ref{fig12}.  

When  the system is weakly coupled to the environment ($\eta = 0.01 \; \eta_{c}$), it exhibits a Markovian decay.  Hence we observe 
a characteristic behaviour where the quantum coherence decreases steadily.  The coherence at time $t=0$ is dependent on the mean photon 
number with quantum coherence being inversely  proportional to the  mean photon number of the system.  The coherence fall decreases with 
increase in the displacement parameter and the squeezing parameter.  Again at higher temperatures, coherence falls faster due to the 
thermal decoherence effects.  In contrast to the Ohmic and the sub-Ohmic case, the system does not exhibit a non-Markovian behavior 
in the strong coupling limit.  Rather we observe a steady decay in the low temperature limit.   In the high temperature limit, initially the coherence 
falls abruptly and then increases slightly and later saturates at a steady value.  The final saturation value depends on the displacement 
and the squeezing parameter.

\section{Steady state analysis}
\label{steadystate}
An important limit of an open quantum system is the long time limit, which can help us to understand steady state behaviour of the 
system.  Towards this end we need the analytic solution of the integro-differential  of $u(t)$ which is given in 
Ref. \cite{zhang2012general} and reads as follows: 
\begin{equation}
u(t) = \mathcal{Z} e^{-i \omega_b t}\!+\!\!\!\int_0^\infty\!\!\!\!\!\!d\omega
\frac{J(\omega) e^{-i\omega t}}{\left[\omega - \omega_0 -
\Delta(\omega) \right]^2 + \gamma^2(\omega)},  \quad
\label{ut}
\end{equation}
The first term arises due to the contribution of the localized mode in the Fano model.  
The second term is due to the continuous part of the spectra which causes dissipation.  
Since the localized mode produces dissipationless dynamics, the system can forever 
memorize some of its initial state information. Here
\begin{equation}
\Delta(\omega) = {\cal P}\int_0^\infty d\omega^{\prime} \frac{J(\omega^{\prime})}{\omega-\omega^{\prime}}
\end{equation}
is a principal value integral and $\gamma(\omega)= \pi J(\omega)$. They are the real and imaginary parts 
of the self-energy correction given by
\begin{eqnarray}
\Sigma(\omega \pm i 0^{+})=\int_0^\infty d\omega^{\prime}
\frac{J(\omega^{\prime})}{\omega-\omega^{\prime} \pm i 0^{+}}=\Delta(\omega) \mp i \gamma(\omega).
\label{sigmaz}
\end{eqnarray}
The conditions for the localized mode frequency is determined by the expression  
$\omega_{b} - \omega_{0} - \Delta (\omega_{b}) = 0$ and 
${\mathcal Z}=\left[ 1 - \Sigma^{\prime}(\omega_b) \right]^{-1}$
is the amplitude of the localized mode. The steady state value of quantum coherence is determined
by the steady state value of the Green's functions given by
\begin{eqnarray}
u(t_s) &=&  {\mathcal Z} \exp(-i \omega_b t_s)  \\
v(t_s) &=&  \int_{0}^{\infty} d\omega [{\mathcal D}_l(\omega) + {\mathcal D}_c(\omega) ]
{\bar n}(\omega,T)
\end{eqnarray}
where ${\mathcal D}_c(\omega)=J(\omega)/[ \left( \omega - \omega_0 - \Delta(\omega) \right)^2 + \gamma^{2}(\omega) ]$ and 
${\mathcal D}_l(\omega)=J(\omega){\mathcal Z}^2/(\omega-\omega_b)^2$.  Using these results we analyze the quantum coherence
steady state values of the system in the strong coupling limit.  We do not investigate the weak coupling limit since the coherence
vanishes in the long time limit for this case.

\begin{figure}
\includegraphics[width=\columnwidth]{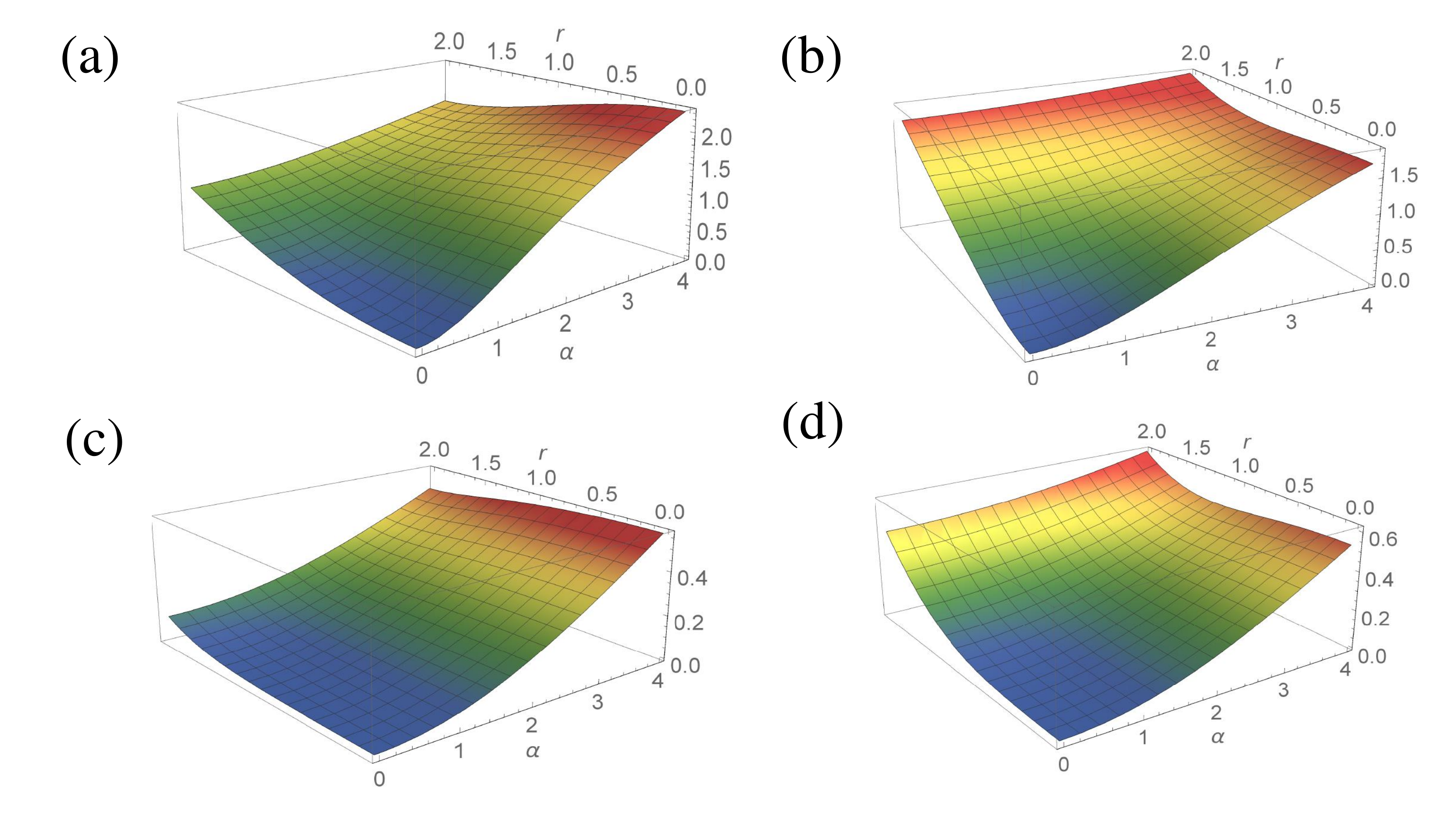}
\caption{The steady state value of coherence in the long time limit $(t \rightarrow \infty)$ for a strongly coupled
system $(\eta = 2.0 \eta_{c})$ in contact with a Ohmic bath. Here the low temperature limit $(T_{s} = 0.1)$
for mean photon number $\bar{n} = 0.1$ and $\bar{n} = 2.0$ is given through the plots in (a) and (b) 
respectively.  The high temperature limit $(T_{s} = 20.0)$ for mean photon number $\bar{n} = 0.1$ and 
$\bar{n} = 2.0$ is given through the plots in (c) and (d) respectively. The cut-off frequency used is 
$\omega = 5.0 \omega_{0}$.}
\label{fig13}
\end{figure}

\begin{figure}
\includegraphics[width=\columnwidth]{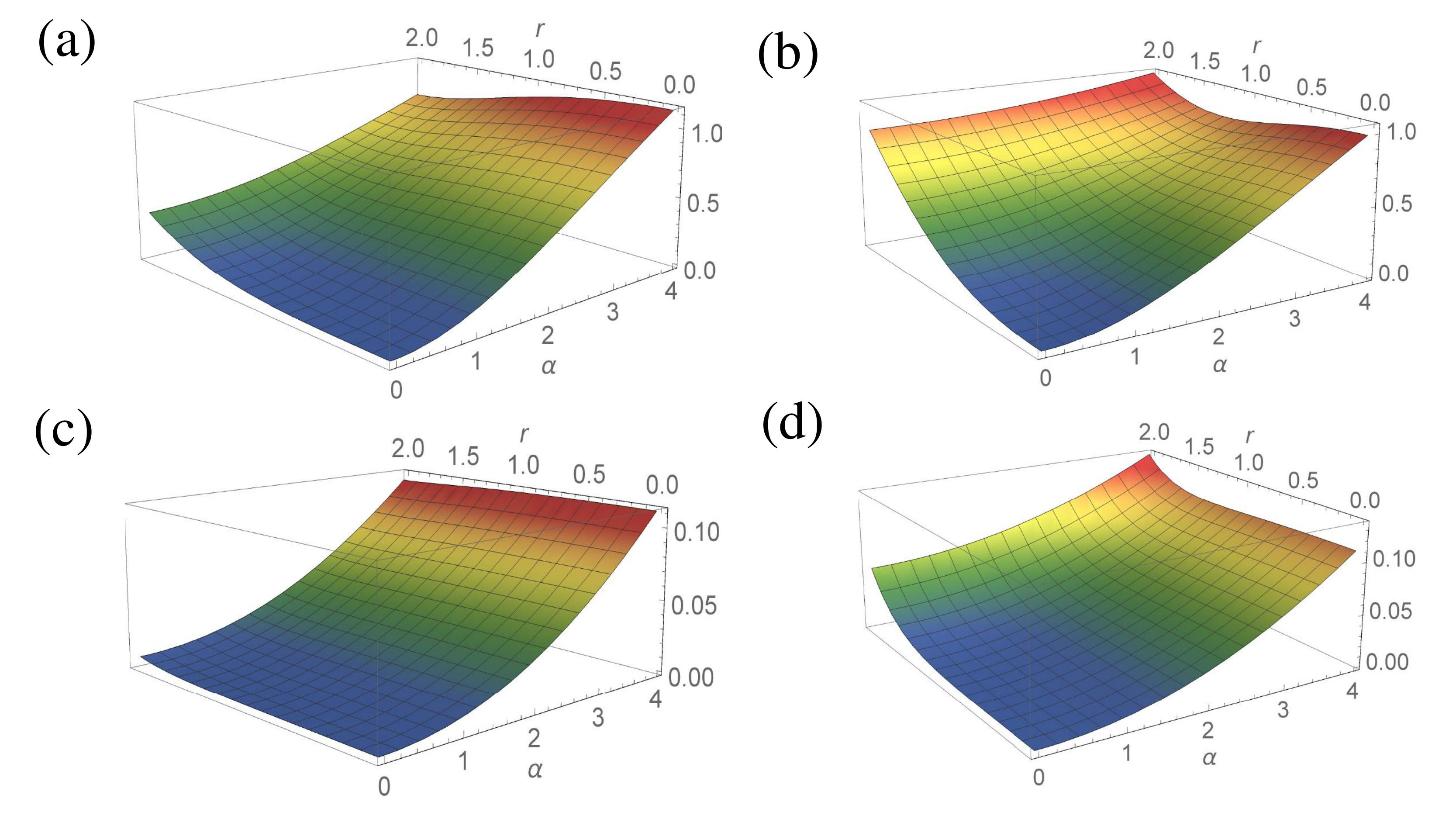}
\caption{In the long time limit $(t \rightarrow \infty)$ for a strongly coupled system $(\eta = 2.0 \eta_{c})$ 
we investigate the steady state value of coherence in contact with a sub-Ohmic bath in this Figure. The low temperature limit 
$(T_{s} = 0.1)$ for mean photon number $\bar{n} = 0.1$ and $\bar{n} = 2.0$ is given through the plots 
(a) and (b) respectively. In plots (c) and (d) the high temperature limit $(T_{s} = 20.0)$ for mean photon number 
$\bar{n} = 0.1$ and $\bar{n} = 2.0$ respectively.  We use a cut-off frequency of $\omega = 5.0 \omega_{0}$.}
\label{fig14}
\end{figure}

\begin{figure}
\includegraphics[width=\columnwidth]{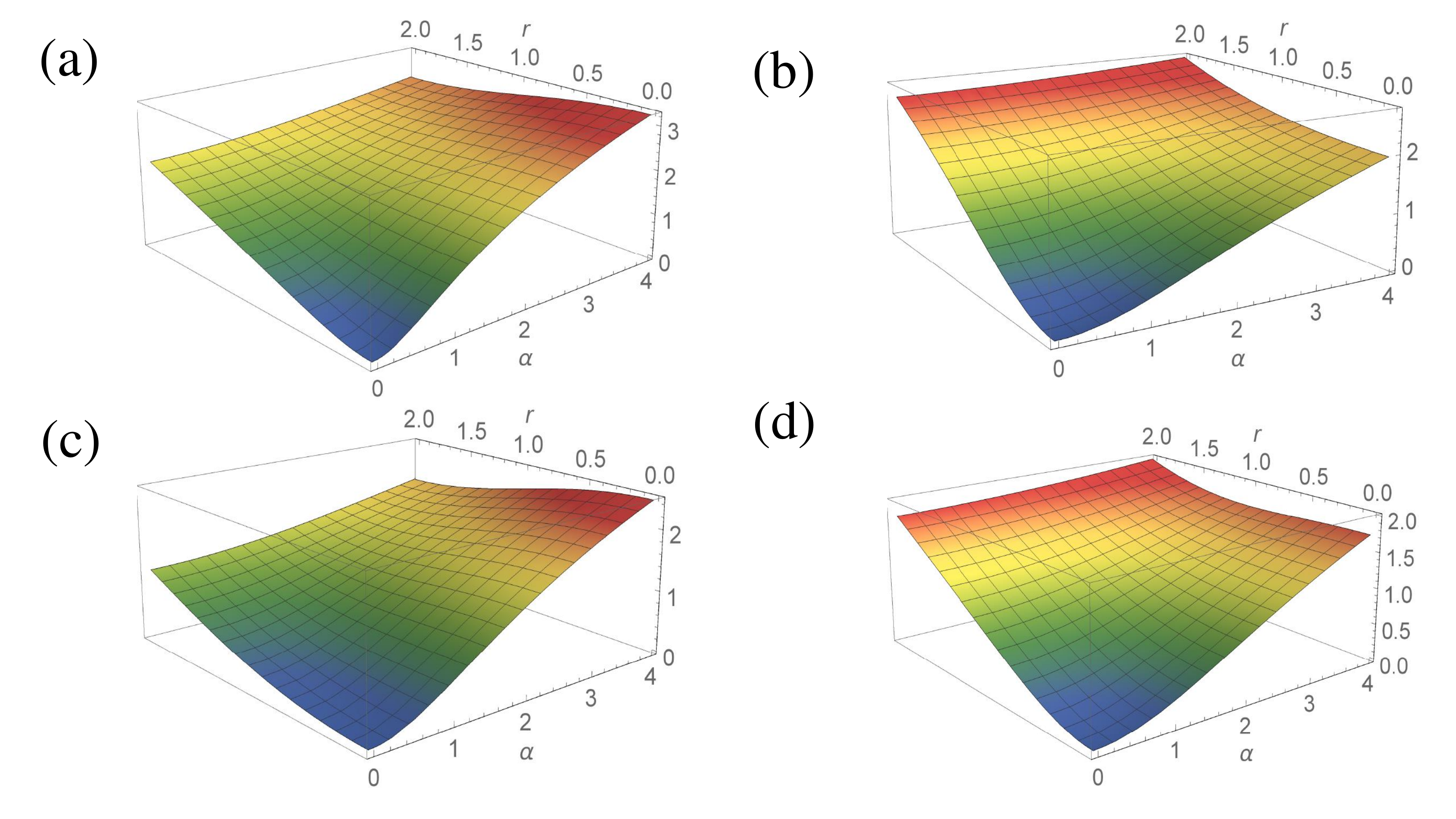}
\caption{For a strongly coupled system, the steady state coherence in contact with a super-Ohmic bath is given 
through the Figure above.  The low temperature limit $(T_{s} = 0.1)$ for mean photon number $\bar{n} = 0.1$ and 
$\bar{n} = 2.0$ is given through the plots (a) and (b) respectively. The high temperature limit $(T_{s} = 20.0)$ 
of the system for the mean photon number $\bar{n} = 0.1$ and $\bar{n} = 2.0$ is given in (c) and (d) repectively. 
Here the cut-off frequency used is $\omega = 5.0 \omega_{0}$.}
\label{fig15}
\end{figure}

\noindent{\it Ohmic case:}
The steady state of quantum coherence for a Ohmic bath with spectral density $J(\omega) = \eta \omega \exp(-\omega/\omega_{c})$ 
is shown in Fig.  \ref{fig13}, where we  show the variation of coherence with the displacement parameter ($\alpha$) and the squeezing parameter ($r$).  
The plots \ref{fig13} (a) and \ref{fig13}(b) show the coherence change for the mean photon numbers of $\bar{n} = 0.1$ and $\bar{n} = 2.0$ 
in the low temperature limit of the environment. In the long time limit, when both the displacement parameter ($\alpha$) and the squeezing parameter ($r$) are 
zero, there is no coherence in the  system.  But if any one of them is finite, there is a finite amount of coherence.   Also we find that the coherence 
varies faster with the decrease in the displacement parameter ($\alpha$) when compared to the squeezing parameter ($r$).   From a comparison 
between the plots \ref{fig13} (a) and \ref{fig13}(b) we find that for low mean photon number $\bar{n} = 0.1$ the coherence decreases with 
increase in the squeezing  parameter whereas  for the high mean photon number $\bar{n} = 2.0$ the coherence increases with the 
squeezing parameter. From the results in \ref{fig13}(c) and \ref{fig13}(d) we find that the long time coherence exhibits the same qualitative 
behavior. But quantitatively, the amount of coherence is lower at high temperatures due to the thermal decoherence effects.

\noindent{\it sub-Ohmic case:}
The coherence variation with the displacement parameter ($\alpha$) and the squeezing parameter ($r$) in the long time limit is 
described for the sub-Ohmic bath with spectral density $J(\omega) = \eta \sqrt{\omega \omega_{c}} \exp(-\omega/\omega_{c})$ through 
the plots in Fig. \ref{fig14}.    The low temperature limit of the coherence variation is analyzed through the plots in \ref{fig14} (a) and \ref{fig14}(b)
where we consider the mean photon number to be  $\bar{n} = 0.1$ and $\bar{n} = 2.0$ respectively.  Here we find that when both the 
displacement and squeezing parameters are zero the coherence vanishes since in this limit the state is an incoherent state.   Comparatively 
the coherence falls faster with the displacement parameter when compared with the squeezing parameter.  The high temperature regime is 
studied through the results in plots \ref{fig14}(c) and \ref{fig14}(d) for $\bar{n} = 0.1$ and $\bar{n} = 2.0$  respectively.   The long time 
coherence in the high temperature limit exhibits the same qualitative behavior as the one in the low temperature limit.  

\noindent{\it super-Ohmic case:}
The super-Ohmic case with spectral density $J(\omega) =\eta  (\omega/ \omega_{c})^{3} \exp(-\omega/\omega_{c})$ is studied for the 
displacement parameter ($\alpha$) and squeezing parameter ($r$) in the long time limit and the results are shown through the plots in 
\ref{fig15}.  For the mean photon numbers $\bar{n} = 0.1$ and $\bar{n} = 2.0$ the low temperature plots are shown through the figures 
in \ref{fig15} (c) and \ref{fig15} (d) respectively.  As expected the coherence is zero when the state is incoherent, i.e., when both the 
displacement and the squeezing parameters are zero.  On increasing the parameters the coherence increases and the rate of increase is 
higher for the displacement parameter than the squeezing parameter.  The results in the  low temperature limit maps similarly to the 
high temperature limit as well.  This can be observed from the study of the results in the plots  \ref{fig15}(c) and \ref{fig15}(d) 
respectively.  The long time coherence in the high temperature limit exhibits the same qualitative behavior as the one in the low temperature 
limit.  But quantitatively the coherence is lower for high temperature values than that of low temperature values.  This is 
because the system exhibits a thermal decoherence at higher values of temperature.

\section{Conclusion}
\label{conclusions}
The quantum coherence dynamics of a single mode squeezed displaced thermal state is analyzed in the present work.  We adopt an 
open quantum system approach where we consider the environment to be a  collection of infinite number of bosonic modes with 
varying frequencies.  The coherence dynamics is studied in the finite temperature limit and considering different values for the 
system-bath coupling strength.  The coherence is measured using the relative entropy measure where the distance to the 
closest thermal state is used.  The number operator of the continuous variable state and the determinant of the covariance matrix 
completely characterizes the quantum coherence.  The  time evolved covariance matrix elements is obtained by solving the quantum 
Langevin equation for the bosonic mode operators.  The two basic nonequilibrium Green's functions namely $u(t)$ and $v(t)$ are 
determined by the time evolution of the field operators. The entire analysis is carried out under three different environmental spectral densities {\it viz} 
Ohmic, sub-Ohmic and super-Ohmic densities.  In the weak interaction limit, when the bath and the system are weakly coupled with 
each other, the quantum coherence decreases monotonically with time.  Meanwhile in the strong interaction limit, we observe 
that the coherence initially decreases but then it increases mildly and shows an oscillatory behavior.  This oscillatory nature, 
signals the presence of non-Markovian behavior and is a characteristic feature of a strongly coupled system.  From our analysis
we find that the system with lower mean photon number has higher amount of initial coherence which falls faster.  Apart from 
investigating the dynamics in a finite time interval, we also look at the coherence evolution in the long time ($t \rightarrow \infty$)
limit.  In this steady state limit, we show the dynamical variation of quantum coherence of the system when it is strongly 
coupled to the environment.  We find that the qualitative behavior of the quantum coherence evolution in the steady state limit 
is same both in the low and high temperature limits.  But quantitatively it is different because of thermal decoherence.  In the present
work we have been able to characterize a mixed Gaussian state completely.  A extension of the study of coherence dynamics to 
non-Gaussian will be the focus of our future works.  Such investigations will require methods beyond the covariance matrix 
approach which works only for Gaussian systems.  

\section*{Acknowledgements}
Md.~Manirul Ali was supported by the Centre for Quantum Science and Technology, Chennai Institute of Technology, India, vide funding number
CIT/CQST/2021/RD-007. R. Chandrashekar was supported in part by a seed grant from IIT Madras to the Centre for Quantum Information, 
Communication and Computing.

\newpage

\section*{References}
\bibliographystyle{iopart-num}
\bibliography{main}

\end{document}